\documentclass[12pt,preprint]{aastex}
\usepackage{graphicx}

\slugcomment{accepted for publication ApJL}

\shorttitle{Integral Field Spectroscopy of Faint Haloes of Planetary Nebulae}
\shortauthors{Monreal-Ibero et al.}

\begin{document}

\title{Integral Field Spectroscopy of Faint Haloes of Planetary Nebulae}

\author{A. Monreal-Ibero\altaffilmark{1}, M.~M. Roth\altaffilmark{1,2}, 
        D. Sch\"onberner\altaffilmark{1}, M. Steffen\altaffilmark{1}, P. B\"ohm\altaffilmark{1}}

\email{amonreal@aip.de}

\begin{abstract}
   We present the first integral field spectroscopy observations of the two
   planetary 
   nebulae 
   \object{NGC~3242} and \object{NGC~4361} with the VIMOS
   instrument attached to VLT-UT3.  By co-adding a large number of spaxels
   we reach an emission line detection limit of
   $5\times10^{-18}$ erg~cm$^{-2}$~s$^{-1}$~arcsec$^{-2}$. In the case of
   \object{NGC~3242}, 
   we succeed in determining some properties of the halo.  The radial
   surface brightness profile in \mbox{[O\,\textsc{iii}]} implies increasing
   mass loss before the formation of the PN. Traces of the mysterious
   `rings' are clearly visible.
   We find for the first time an apparent temperature gradient across a halo:
   from about 16\,000~K close to the shell/halo transition to 20\,000~K at the
   halo's outer edge.  No line emission is seen in the suspected
   halo region of \object{NGC~4361} down to the sensitivity limit.

\end{abstract}

\keywords{planetary nebulae: individual (\object{NGC 3242},
\object{NGC 4361})}


\altaffiltext{1}{Astrophysikalisches Institut Potsdam, An der Sternwarte 16,
  D-14482  Potsdam, Germany}
\altaffiltext{2}{Visiting astronomer at the European Southern Observatory, Chile (ESO Proposal No. 073.D-0576)}

\section{Introduction}
\label{INTRODUCTION}

Next to supernova explosions, stellar winds on the asymptotic
giant branch (AGB) make significant contributions to the recycling of
chemically enriched material to the interstellar medium. The physics
of AGB mass loss is therefore a key ingredient for our understanding of
the evolution of galaxies, as far as the recycling of matter, processed by
different stellar populations, is concerned.  Despite considerable progress in 
our theoretical understanding of stellar evolution and mass loss during the
last AGB phase, 
observational details of this critical period have remained obscure.
Haloes of planetary nebula (PN) are fossil records of the mass loss history
at the tip of the AGB, which, in principle, can be investigated with the
plasmadiagnostic tools for gaseous nebulae. However, spectroscopy of these 
low surface brightness regions is extremely difficult to perform with 
conventional slit spectrographs.

Encouraged by successful experiments with the PMAS instrument
\citep{rot04,rot05}, we have employed the technique of integral field
spectroscopy (IFS) to obtain unprecedented sensitivity for the measurement of
PN emission line intensities, by co-adding very many spatial elements
(spaxels) over the field-of-view (FOV) of an integral field unit (IFU).
We have used the unique light-collecting properties of the VIMOS IFU
at the VLT for a plasma diagnostic analysis of selected targets
from the catalogue of PN haloes of \citet{cor03} with the
goal to measure $n_{\rm e}$, $T_{\rm e}$, and chemical composition in order to
test the most recent theoretical predictions.

Here we present first results from the spatially resolved spectrophotometry in
the outskirts of NGC~4361, which is a low-metallicity galactic halo object, and
of the disk PN NGC~3242. Both objects are well developed PNe with
central stars of very similar effective temperatures, viz.\ of $75\,000~{\rm
  K}$ for \object{NGC 3242} and $82\,000~{\rm K}$ for \object{NGC 4361},
  respectively \citep{men92}.
  \object{NGC 3242} has a well defined  halo at the
  percent level of the peak surface brightness \citep{cor03}, where recently
  mysterious `rings' have been detected \citep{cor04}.  To date, no halo has
  been reported for 
  \object{NGC 4361}.  Also, due to its low metal content, the electron
  temperature of \object{NGC 4361} ($\simeq 19\,000~{\rm K}$) is unusually
  large (Torres-Peimbert et al.\ 1990).

\section{Observations}
\label{OBSERVATIONS}

We performed VLT observations at UT3 ``Melipal'' on April 17 and 18, 2004, 
using the VIMOS-IFU.
The instrument was setup in the LR\_blue mode with a spaxel scale of
0\farcs67, a FOV of $54^{\prime\prime} \times54^{\prime\prime}$,
a nominal wavelength range 
of 3700--6700~{\AA}, 5.3~{\AA}~pix$^{-1}$ reciprocal dispersion, and a
spectral resolution of $R\approx180$. Both nights were photometric,
with a seeing of $\approx1\farcs3$ and $\approx1\farcs0$
FWHM, respectively. We observed each object with a
series of snapshot exposures centered on the central star in order to
obtain a reference in the bright part of the nebula, and a series of
deep exposures in one or two halo fields as the major objective of
this run: $3\times 100$~s centered on NGC3242, $3\times 900$~s 
offset by 55 and 63 arcsec to the west (halo fields), 
$2\times2$ mosaic centered on NGC~4361 with 300~s each, and
$2\times 900$~s adjacent to this mosaic to the west 
(see Fig.~\ref{NGC3242MAP}, Fig.~\ref{NGC4361MAP}).
Internal continuum and arc
lamp flatfield as well as spectrophotometric
standard star exposures were taken throughout the night.


\section{Data Reduction}

The data reduction was performed with a modified subset of routines
from the PMAS P3d pipeline \citep{bec02,rot05}. 
An input list of data and calibration files was processed by an IDL 
script such that each of the four VIMOS channels was treated
separately, dissecting each CCD frame into four subfields which
correspond to a bank of spectra from the same pseudo-slit.
Thus a total of 16 subfields was piped into the P3d routines, which
then operated like with ordinary PMAS frames.

The data reduction
proceeded as follows: firstly, the bias level was subtracted and
cosmic ray hits were removed. Secondly, a trace mask was generated
from an internal continuum calibration lamp exposure, identifying
the location of each spectrum on the CCD along the direction of
dispersion. In order to make this procedure robust, it was necessary
to truncate the spectra on each side and discard regions with
contamination from adjacent spectra of a neighbouring bank. Since VIMOS
calibration spectra are taken only when the telescope is pointing
to the zenith, and because of the presence of significant flexure,
the trace mask had to be shifted in x and y to match the location
of the actual science exposure. The offset for this task was obtained 
from a cross-correlation between the calibration lamp and science
exposure.
Thirdly, the $6400$ spectra were extracted using a simple swath
extraction technique (no profile fitting). After this, the data had
changed from a CCD-based format to the so-called row-stacked-spectra
(RSS) format, which is a 2-dimensional image where each row represents
a spectrum. As the fourth step, the RSS frame was wavelength-calibrated
with an arc exposure, which turned out to be a critical step since
spectral line artefacts (zero and higher order contamination)
sometimes confused the P3d line search algorithm. The last step
consisted in a correction of the spectrum-to-spectrum sensitivity
variation, which was performed by dividing by an extracted and
normalized continuum lamp flatfield exposure.

Flux calibration was performed from a series of standard star
exposures which were taken in different quadrants of the IFU
in order to obtain an idea of the achievable accuracy. 
There is considerable scatter, indicating that the flux calibration 
is not uniform over the face of the IFU. Secondly, an inspection
of the standard deviation as a function of wavelength shows
that only within a window of [4200~{\AA},6100~{\AA}] the
variation is reasonably behaved ($\approx15$~\% r.m.s).
The region of the spectra beyond 6100~{\AA} presents an artifact probably due
to other orders contamination. In addition, wavelength
calibration in this range was not accurately enough and some spectra
lacked a red continuum for H$\alpha$ which prevents us for using this line in
the analysis. 



\section{Results and Discussion}
\label{RESULTS}

\subsection{Maps, IFU defects, PN morphologies}

\begin{figure}[h!]
\epsscale{.80}
\includegraphics[width=\linewidth,angle=0, clip=,bbllx=10, bblly=5,
bburx=580, bbury=565]
{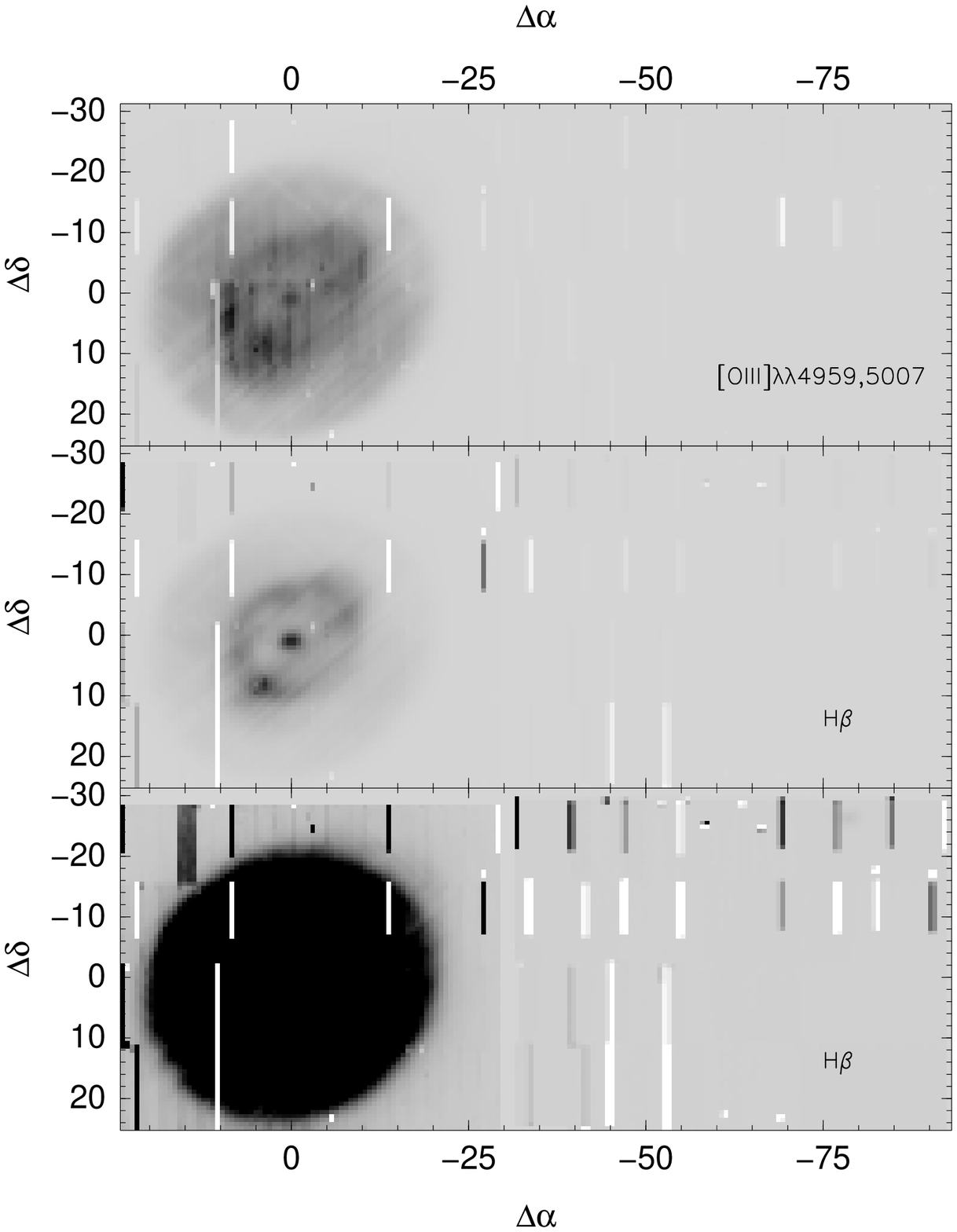}
\caption{Maps of NGC~3242 in
  \textsc{[O$\;$iii]}$\lambda\lambda$4959,5007~{\AA} (top), H$\beta$ low 
contrast (middle), and high contrast (bottom). Mosaic constructed from
shallow pointing, centered on the PN, and two overlapping deep
pointings to the west. Artifacts are discussed in text. 
Orientation: N up, E left.\label{NGC3242MAP}}
\end{figure}

\begin{figure}[h!]
\epsscale{.80}
\includegraphics[width=\linewidth,angle=0, clip=,bbllx=10, bblly=5,
bburx=580, bbury=565]
{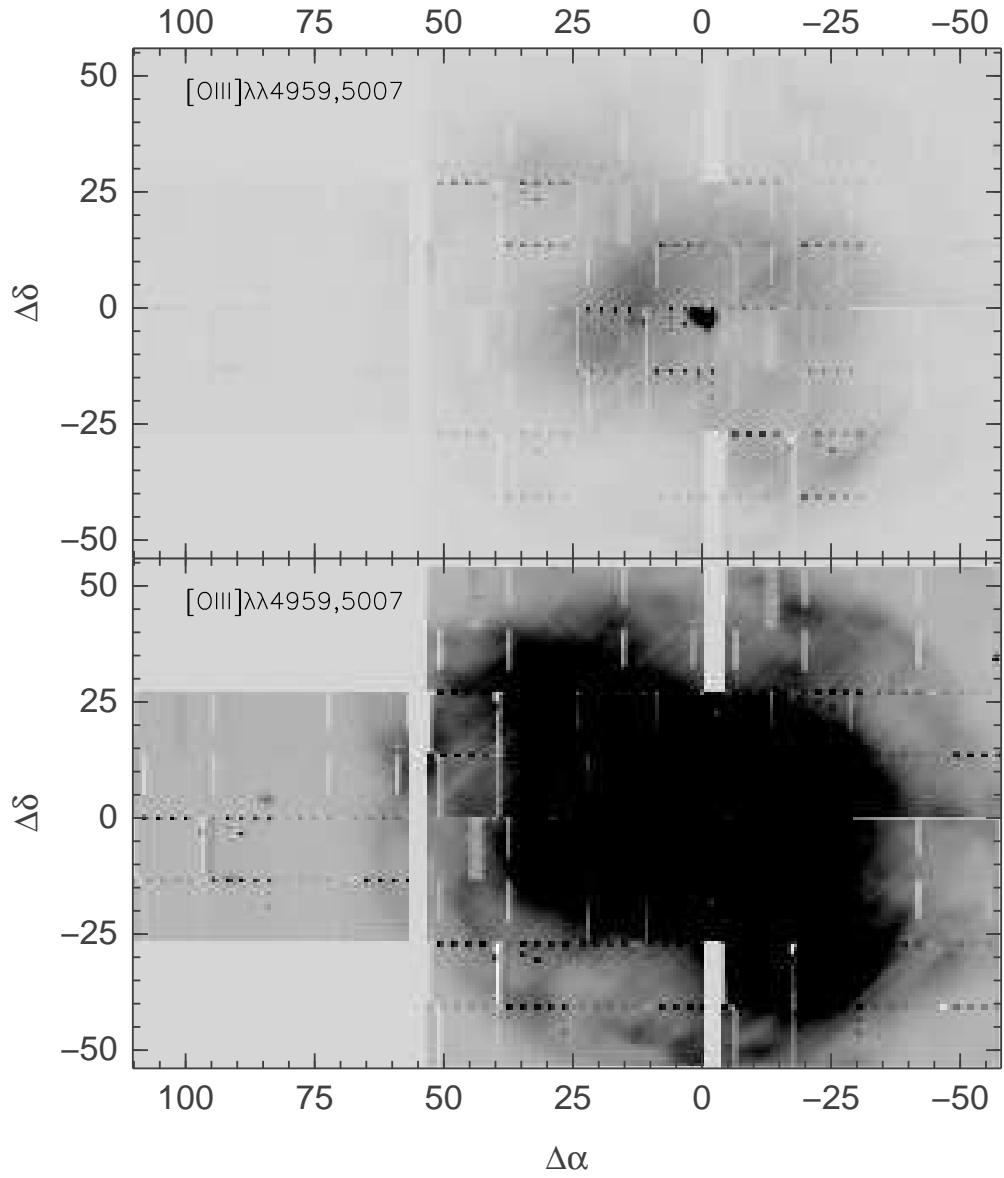}
\caption{Maps of NGC~4361 in
  \textsc{[O$\;$iii]}$\lambda\lambda$4959,5007{\AA}. Mosaic constructed
  from 5 shallow pointings centered on the PN, and one deep pointing
  offset to the east. Low contrast (top), high contrast (bottom).
  Orientation: N up, E left.\label{NGC4361MAP}}
\end{figure}

Figures~\ref{NGC3242MAP} and \ref{NGC4361MAP} show maps of \object{NGC~3242}
and 
\object{NGC~4361} in the emission lines of
\textsc{[O$\;$iii]}$\lambda\lambda$4959,5007~{\AA} and H$\beta$,
respectively.
Various IFU defects are scattered over the FOV as distinct rectangles, which 
needed to be discarded from the subsequent analysis.
The limited accuracy of the calibration of spaxel-to-spaxel response variation 
is visible in some vertical or diagonal intensity enhancements, e.g.\ in the
\textsc{[O$\;$iii]}$\lambda\lambda$4959,5007~{\AA} and H$\beta$ frames of
\object{NGC 3242}. 

Despite these cosmetic limitations, the fundamental morphological appearance
as described in \citet{cor03} is clearly visible as far as the
brighter parts of the nebulae are concerned: central star, central cavity with
enhanced rim, and shell.  This gross morphology can be fully explained by the 
combined action of thermal pressure due to the heating of the nebular gas by
photo-ionization and of dynamical pressure excerted by the fast stellar
wind \citep[cf.][]{per04}.  As can be seen from the
Figs.~\ref{NGC3242MAP} and \ref{NGC4361MAP}, the surface-brightness
distribution of \object{NGC 4361} appears to be much smoother, with no clear
distinction between rim and shell.  This difference in morphology may well be
caused by the low metallicity, since it is known that stellar winds become
less vigorous for metal-poor stars \citep[cf.][]{vin01}.
A faint halo is not directly visible in
any of the maps at the chosen constrast level.


\subsection{Emission line intensities, radial profiles}

\begin{figure}[h!]
\includegraphics[width=\linewidth, angle=0, clip=,bbllx=00, bblly=270,
bburx=600, bbury=605]
{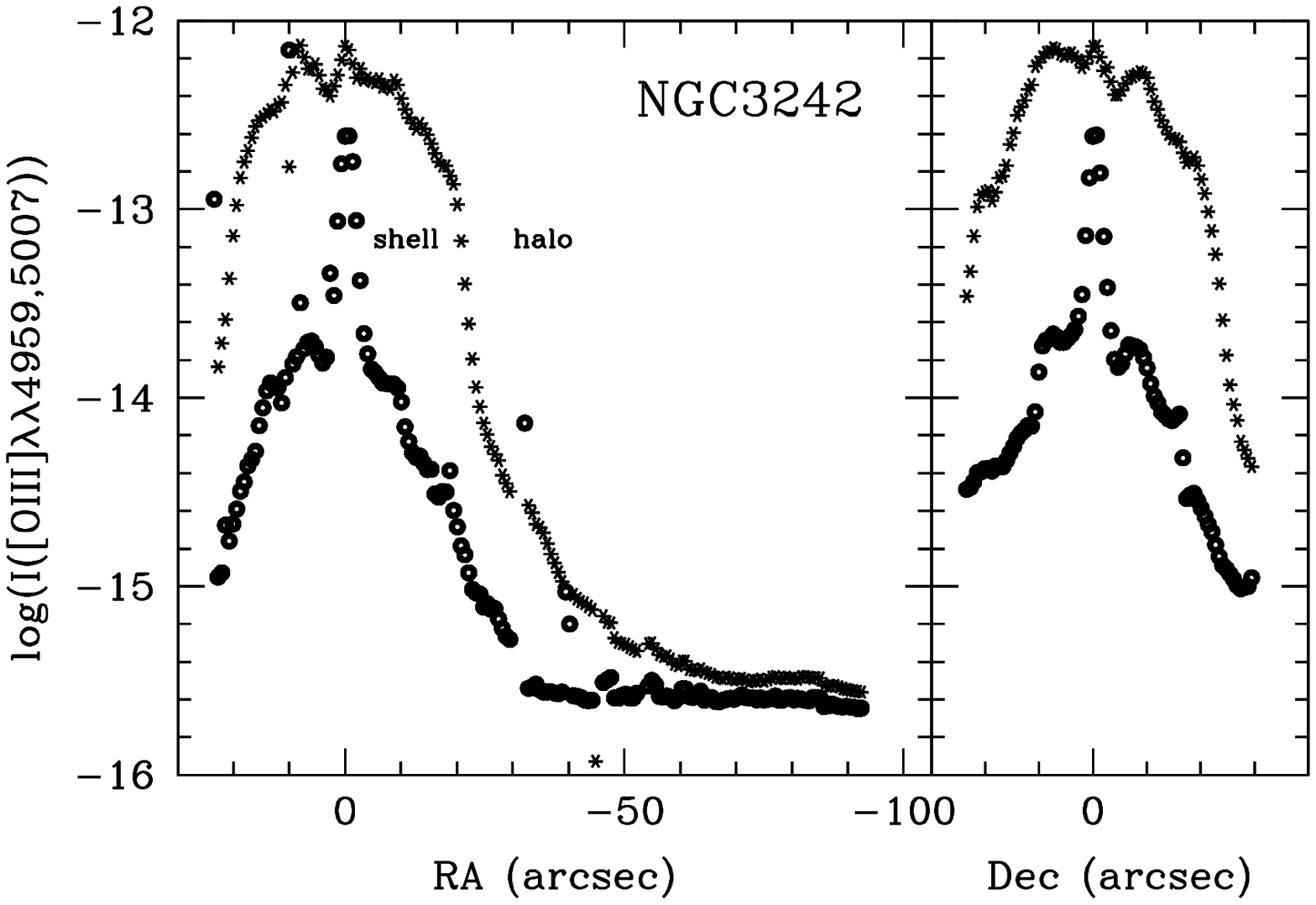}\\
\includegraphics[width=\linewidth, angle=0, clip=,bbllx=00, bblly=270,
bburx=600, bbury=605]
{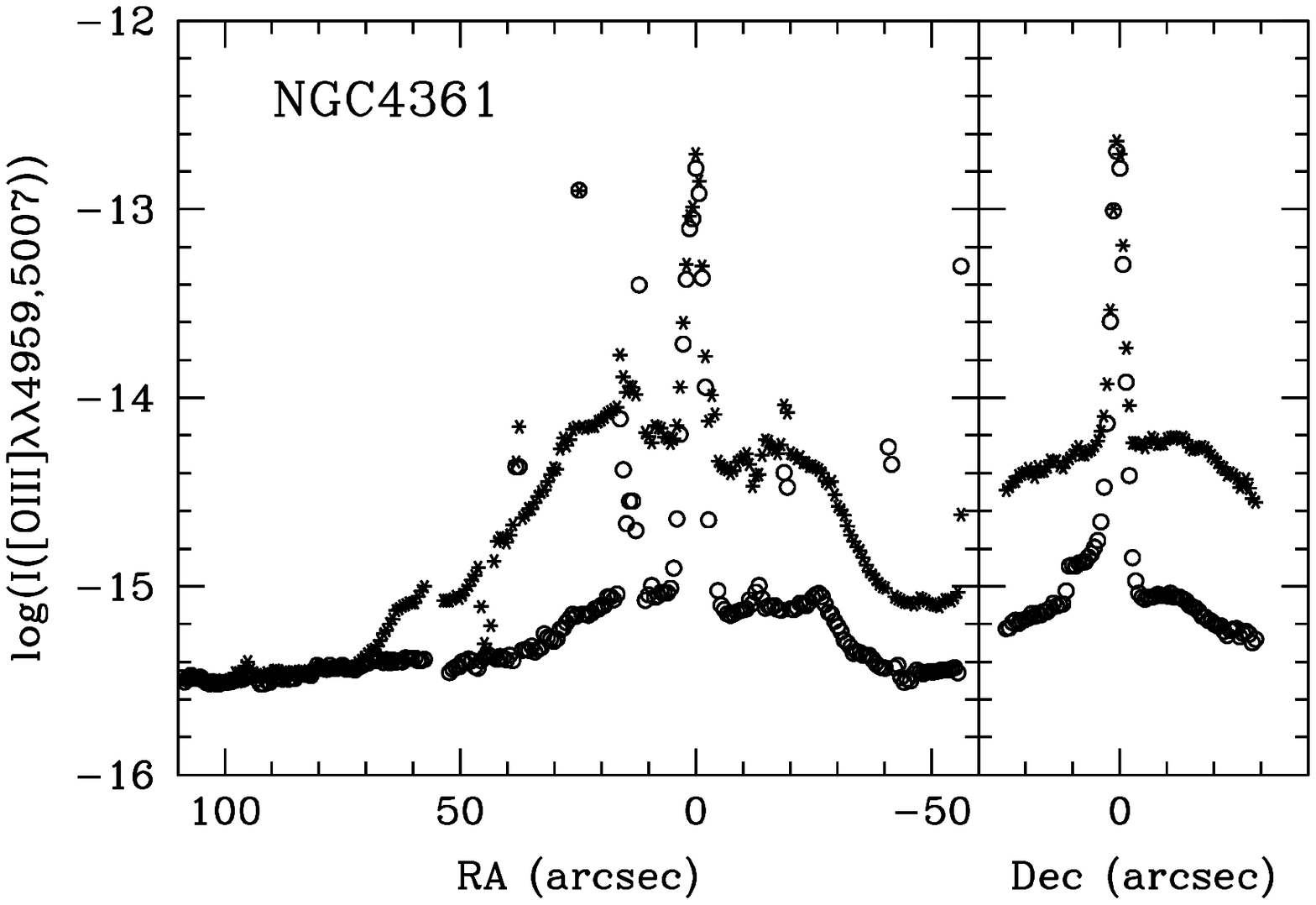}\\
\caption{Radial intensity profiles in [O\,\textsc{iii}] + background
  (asterisks) and background (circles) for NGC~3242 (top), and NGC~4361
  (bottom), from central cuts one spaxel wide and along right ascension and
  declination. The \textsc{[O\,iii]}$\lambda\lambda$4959, 5007~{\AA} emission
  line intensity is integrated over the interval 4900--5068~\AA, the
  background is determined from the interval 5100-5150~{\AA} and corrected for
  the different band width. The intensity scale is is in
  erg~cm$^{-2}$~s$^{-1}$~arcsec$^{-2}$. 
\label{PROFILES}}
\end{figure}

Fig.~\ref{PROFILES} shows observed and continuum substracted radial profiles
in  \textsc{[O$\;$iii]}$\lambda\lambda$4959, 5007~{\AA} along
RA and DEC axes.
The faint nebular continuum and emission lines make an
appreciable background contribution, which is
roughly a factor of 50 below the \textsc{[O$\;$iii]} lines intensity, but
which becomes negligible as the former approaches the level of the night sky
continuum background. For \object{NGC~3242}, this happens beyond a radial 
distance of $\approx40$~arcsec 
from the central star, from where we have chosen an average as an estimate
of the true sky background at the wavelength of the \textsc{[O$\;$iii]} lines
(\object{NGC~4361}: also beyond $\approx40$~arcsec). Using this procedure, we
measure the averaged halo \textsc{[O$\;$iii]}$\lambda\lambda$4959,5007~{\AA}
 intensity, corrected for sky, as  
4.1$\times 10^{-16}$~erg\,cm$^{-2}$\,s$^{-1}$\,arcsec$^{-2}$, which is a
factor of $\approx2000$ below the the intensity of the rim.
For \object{NGC~4361}, having a
more than two orders of magnitude fainter surface brightness than
\object{NGC~3242}, no halo emission is seen in Fig.~\ref{PROFILES}.

\begin{figure}[h!]
\epsscale{.80}
\includegraphics[bb= 20 270 590 605, width=\linewidth,angle=0]
{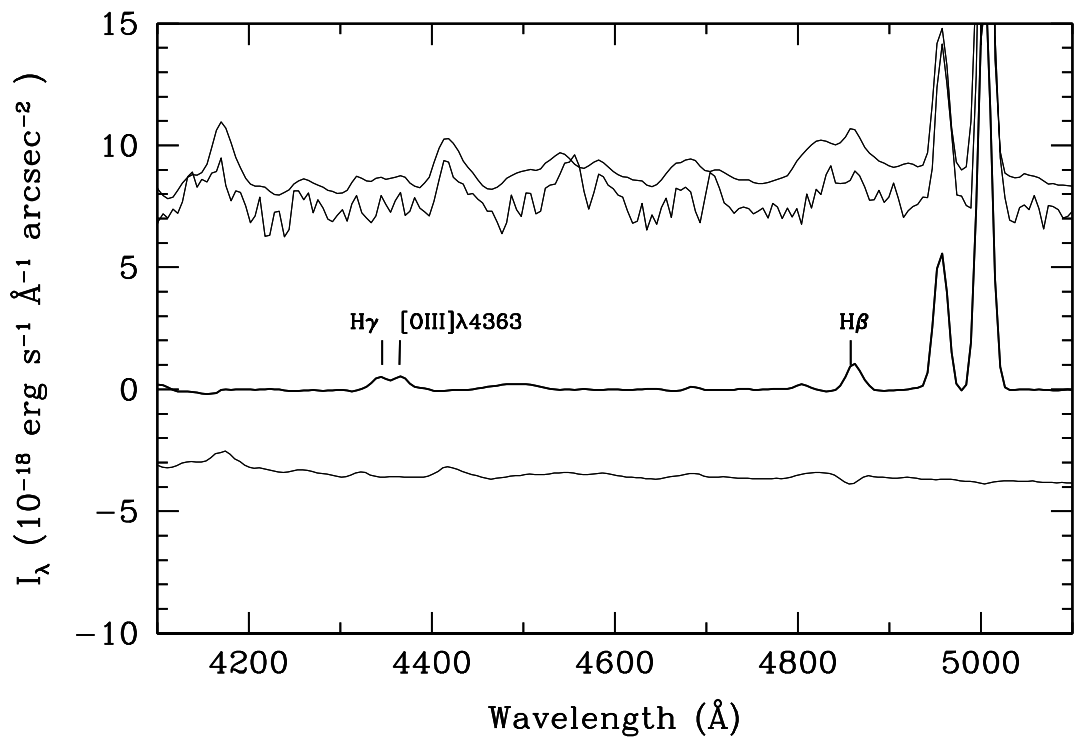}
\caption{Mean halo spectra of NGC~3242 and NGC~4361, averaged from
  a total of 4547 and 4312 spectra, resp.,  and plotted in units of
 $10^{-18}$~erg~cm$^{-2}$~s$^{-1}$~\AA$^{-1}$~arcsec$^{-2}$.
From top to bottom: NGC~3242 co-added (shifted by
 $+1\times10^{-18}~\mbox{erg\,~cm$^{-2}$\,s$^{-1}$\,\AA$^{-1}$\,arcsec$^{-2}$}$),   
NGC~3242 single-spaxel, NGC~3242 co-added + sky-subtracted, NGC~4361 co-added
 + sky-subtracted 
spectrum (shifted by $-5\times10^{-18}~\mbox{erg\,cm$^{-2}$\,s$^{-1}$\,\AA$^{-1}$\,arcsec$^{-2}$}$).
  \label{HALOSPECTRA}}
\end{figure}

In order to increase the sensitivity for the outer halo regions, we
co-added very many spaxels
over extended regions beyond the shell to obtain an estimate of the
average emission line intensity of the halo. Fig.~\ref{HALOSPECTRA}
illustrates the results.  For \object{NGC~3242}, the
average spectrum clearly confirms the presence of an emission-line halo.
Besides the detection of H$\beta$,
which is blended with a sky background feature, there is also the
line blend of H$\gamma$ and \textsc{[O$\;$iii]}$\lambda$4363, the latter
 being important for measuring the electron temperature.
We have measured mean intensities for H$\gamma$, [O\,{\sc iii}]$\lambda$4363,
H$\beta$ and [O\,{\sc iii}]$\lambda\lambda$4959,5007 of
(8.9, 9.9, 19.8, 411.9) and (51.4, 66.8, 114.7, 4426.4)$\times
10^{-18}$~erg\,cm$^{-2}$\,s$^{-1}$\,arcsec$^{-2}$ for the outer and inner
region of the halo respectively.
Obviously,
the accuracy of this result would significantly benefit from higher spectral
resolution, e.g.\ the VIMOS-IFU HR modes, offering
a 10-fold higher resolving power than our current data, at the
expense of a 4 times smaller FOV.

 In the case of \object{NGC~4361}, however, there is no detection of halo line
 emission.   We derived the detection limit by simulating emission lines with 
different intensities, which where superimposed on the co-added
\object{NGC~4361} 
spectrum, and attempting to recover these lines with gaussian fits.
Our detection limit estimate in the halo of \object{NGC~4361}
 is $5\times10^{-18}$~erg\,cm$^{-2}$\,s$^{-1}$\,arcsec$^{-2}$. This value is
 not quite 2 orders of magnitude below the rim surface brightness, i.e.\ 
not sufficient for the typical intensity contrast of $\approx10^3$
 \citep{cor03}. 
Also for this object, suppressing the sky background with higher spectral
resolution would significantly lower the detection limit, most probably down
to the expected halo surface brightness on the order of
$1\times10^{-18}$~erg\,cm$^{-2}$\,s$^{-1}$\,arcsec$^{-2}$.

\subsection{The halo of NGC 3242}

\begin{figure}[h!]                                     
\includegraphics*[bb= 30 270 590 620, width=\linewidth]
{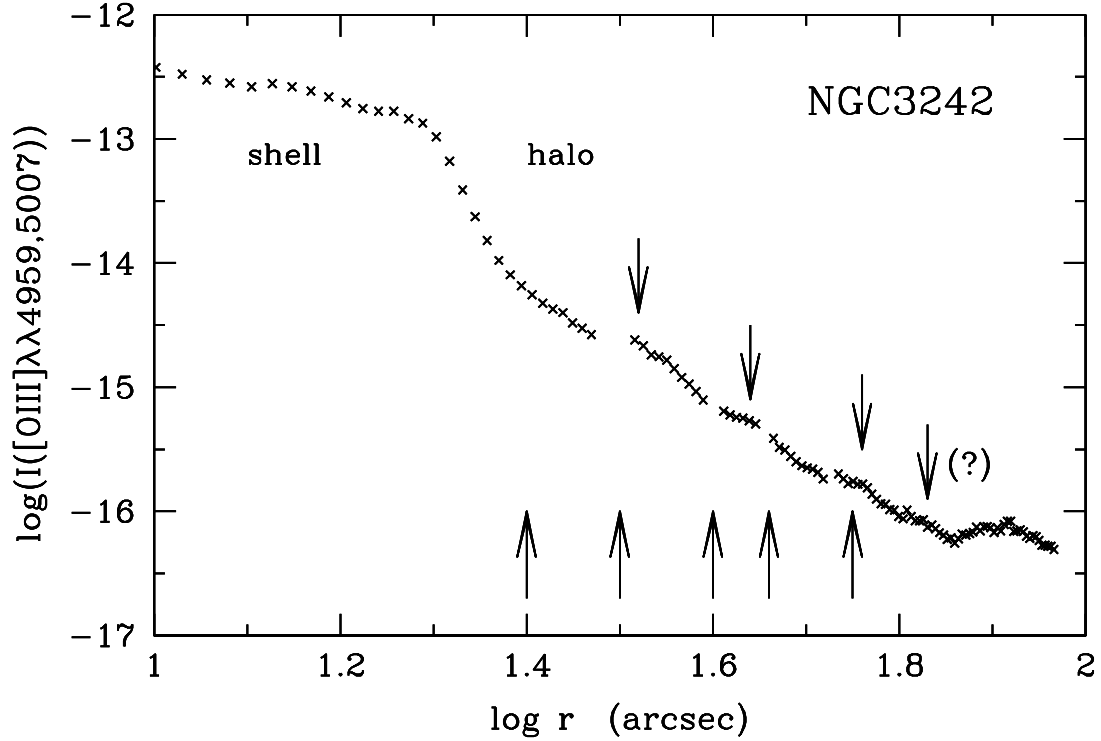}
\caption{Radial background-subtracted \textsc{[O\,iii]}$\lambda\lambda$
  4959,5007~{\AA} intensity profile for NGC~3242, one spaxel wide, along right ascension, in erg~cm$^{-2}$~s$^{-1}$~arcsec$^{-2}$.
The shell-halo transition  is at a distance of
$\simeq 22$~arcsec from the center.   The vertical arrows indicate the
position of the `rings' from our cut (\emph{downwards}) and from Corradi
et al.\ (2004, priv.\ comm.) (\emph{upwards}).  The gaps are due to the defects
of the IFU.
}
\label{HALO.3242}
\end{figure}

The background-subtracted intensity profile in
\textsc{[O\,iii]}$\lambda\lambda$4959,5007~{\AA} for the halo of
\object{NGC~3242} is shown in
Fig.~\ref{HALO.3242}.  We attribute small 
intensity `bumps' to the existence of so-called halo `rings' found recently
in a number of PNe \citep{bal01,cor04}.
The agreement with the radial positions given by Corradi et al.\ is good,
except in a few cases.  We have no indication of a `ring' at
25$^{\prime\prime}$, and 
instead of the 2 `rings' at 40 and 46$^{\prime\prime}$ we found only 1 located
at a 
radial distance of $43^{\prime\prime}$.   Our marginal detection at $\approx
67^{\prime\prime}$
corresponds to an arc visible in Fig.~1 of Corradi et al.\ at position
angle $\sim 0$--$40^{\mathrm{o}}$, but not annotated as `ring' by the authors.

The halo of \object{NGC~3242} is obviously limb-brightened, as the wide bump
around 80$^{\prime\prime}$ indicates. These haloes are very common and are
explained by hydrodynamical effects when a strong AGB wind interacts with
slower, less dense  matter expelled earlier during the aftermath of a helium
shell flash \citep{sch02}.  Matter piles up into a denser 
shell which, once ionized, is resposible for the bright limb of the halo.

  The overall slope of the halo brightness, however, is constant and can well
  be approximated by a power-law representaion, $I\propto r^{-\beta}$, with
  $\beta = 4.5$.  With the reasonable assumption that in the halo all the
  oxygen is doubly ionized one can deduce the radial density profile of the
  halo as a power-law profile $n(r)\propto r^{-2.8}$.   Such a rather strong
  density decline with distance from the star is consistent with the rather
  large shell expansion velocity of 36~km\,s$^{-1}$ and hints to strongly
  increasing mass loss towards the end of the AGB evolution
  \citep[cf.][Fig.~13 therein]{sch05}. 

We determined the electron temperature, $T_\mathrm{e}$, from the
   \textsc{[O\,iii]} lines 
  in the vicinity of the
  main body of the PN and at the outer edge of the halo, employing the
   low-density limit, using a value for the extinction of $c=0.15$
   \citep{bal93} and the extinction curve of \citet{flu94}.  We found 
  an apparent temperature gradient across the halo: close to the PN
  shell (${\rm RA}\ga -25~{\rm arcsec}$, with 766 spaxels) we have
  $T_\mathrm{e} = 15\,700$~K, and at the halo's edge
  (${\rm RA}\la -60~{\rm arcsec}$, with 4547 spaxels),
  $T_\mathrm{e} = 20\,300$~K.  For comparison, we determined also the electron
  temperatures of the rim (i.e.\ the bright part of the PN) and the 
  shell and found, as averages over $5\times5$ spaxels, $T_\mathrm{e}= 12\,000$
  and $T_\mathrm{e} = 11\,000$~K, resp., in good agreement with the
  determinations of Balick et al. (1993).  Our error estimates are
  $+2000,-1200$~K for the halo and $+600,-500$~K for the PN.


   The existence of hot haloes has already
   been reported in the past for \object{NGC~6543}, \object{NGC~6826}, and
   \object{NGC~7662} \citep{midd89, midd91}, based on long-slit observations.
   The (mean) halo temperatures range
   from 13\,000~K to 17\,500~K, but temperature gradients have not been
   established.  High temperatures in the halo can easily be set up by the
   rapid 
   passage of an ionization front once the main PN becomes optically thin for
   ionizing photons \citep{mar93}. Due to the low densities, the halo
   matter is far from being in thermal equilibrium, and a positive radial
   temperature gradient will develop because the cooling rate depends on
   density squared.  Judging from the existence of hot haloes around PNe with
   quite different evolutionary stages, the cooling times seem
   to be comparable with the nebular lifetime, an important
   constraint for hydrodynamic models.


\section{Summary and Conclusions}

IFS has shown to be capable of 
performing spatially resolved spectrophotometry down to sky-limited
intensity levels. We have been able to find halo emission lines
down to a detection limit of
$5\times10^{-18}$~erg~cm$^{-2}$~s$^{-1}$~arcsec$^{-2}$, 
with positive detections of \textsc{[O$\;$iii]}$\lambda\lambda$5007,4959~{\AA},
\textsc{[O$\;$iii]}$\lambda$4363~{\AA}
and H$\beta$ in the halo of \object{NGC~3242}.  We confirmed the existence of
the mysterious `rings' and found, for the first time, an apparent electron
temperature gradient across the halo which, if confirmed,
poses a challenge for evolutionary models of planetary nebulae.  Haloes are
obviously not in thermal equilibrium, making a plasma analysis based on steady
state photoionization models highly questionable.
From the steep radial intensity profile of the halo it follows that the
mass loss rate must have increased until the very end of the AGB evolution.
No emission line is seen in the putative halo region of \object{NGC~4361} down
to the sensitivity limit.  Owing to the low surface brightness of this PN,
the expected halo is too faint to be detected with the present low-resolution
observing mode.

\acknowledgments

AMI acknowledges support from the Euro3D Research Training Network,
funded by the EC (HPRN-CT-2002-00305). Part of this work was supported
by the ULTROS project, funded by the German Verbundforschung
(05AE2BAA/4).


\clearpage

\clearpage

\clearpage

\clearpage

\clearpage

\end{document}